\documentstyle[seceq,preprint]{jpsj}

\def\lesssim{ \raise2pt\hbox{$\mathop{\! <}%
               \limits_{\raise2pt\hbox{$\sim$}}$} }                 
\def\gtrsim{ \raise2pt\hbox{$\mathop{>}%
               \limits_{\raise2pt\hbox{$\sim$}}$} }

\def\runtitle{
Bosonization Approach to the Kondo Lattice 
}
\def\runauthor{S. Fujimoto and N. Kawakami}

\title{Bosonization Approach to the One-Dimensional Kondo Lattice Model
at Half Filling: Reexamination}

\author{Satoshi Fujimoto$^1$ and Norio Kawakami$^2$}
\inst{$^1$Department of Physics, Kyoto University, Kyoto 606, Japan\\
$^2$Department of Applied Physics,
and Department of Material and Life Science, \\
Osaka University, Suita, Osaka 565, Japan}

\recdate{}
  
\abst{
Low-energy properties of the one-dimensional Kondo lattice model are
investigated by using bosonization
and renormalization
group methods. The formation of the spin and charge 
excitation gaps at half-filling 
is discussed in detail both for antiferromagnetic and
ferromagnetic Kondo couplings. Also, 
the effect of the interaction between conduction electrons
on the gap formation is studied both for the Kondo insulating phase 
and for the Haldane gap phase.  
It is found  that while the repulsive interaction 
changes the dependence of 
the excitation gap on the Kondo exchange coupling,
the attractive interaction can drive the Kondo (or Haldane)
insulating phase to
a completely different phase of massless metallic states.
}
\kword
{Kondo lattice, heavy fermion, bosonizaton, spin gap}
\begin{document}
\sloppy
\maketitle

\section{Introduction}

The Kondo lattice model is a fundamental model for heavy electron
systems such as Cerium or Uranium compounds.
One-dimensional (1D) version of the model has been 
studied extensively for its simplicity
by using numerical as well as 
analytic methods.\cite{sca,tsu,tro,ueda,tsvelik,fuji,ek,mou,white}
The model shows various phases according to the electron density and 
the strength of the Kondo coupling.
At half-filling the Kondo insulating phase realizes, which 
has the excitation gaps both
in the spin and charge sectors.\cite{tsu,fuji}
Away from half-filling the model shows the Tomonaga-Luttinger 
liquid phase for small Kondo exchange couplings.
\cite{ueda,fuji,mou} More recently,
the effect of interaction between conduction electrons 
on these phases has been studied numerically.\cite{shi}

We have previously investigated this model by using bosonization 
method\cite{fuji}, which is a powerful tool to study 
low-energy properties of 1D quantum systems.
In this method, one needs to linearize the 
dispersion of electrons at the first stage for
deriving a low-energy effective hamiltonian.  
However, since localized spins have no dispersion, 
it is rather difficult 
to apply this method to the Kondo lattice model
straightforwardly. In the previous paper, to bypass this difficulty 
we naively introduced a finite velocity for
the excitation of localized spins in the 
microscopic model, and then performed bosonization.\cite{fuji} 
This prescription, however, may not give 
desirable functional forms of the spin and charge gaps
for the original Kondo lattice model: the predicted 
gaps are not consistent with
those obtained by numerical studies.\cite{tsu}
In this paper, by starting with the original Kondo lattice model,
we develop a low-energy effective theory improved in this respect,
and reexamine the formation of the 
spin and charge gaps at half filling
both  for antiferromagnetic (AFM) and ferromagnetic (FM)
Kondo couplings. Based upon this formulation, we also
study how the short-range interaction between conduction electrons 
affects the spin/charge gap formation.

The organization of the paper is as follows.
In \S 2 we introduce the bosonized effective hamiltonian for 
the 1D Kondo lattice model, and derive 
the scaling equations for various couplings
which characterize the low-energy fixed point properties.
In \S 3 we first discuss 
the case of the AFM Kondo coupling, and describe
the formation of the spin and charge gaps 
in the Kondo insulating phase.
The effect of interaction between conduction electrons 
on this phase is also investigated.  It is shown that
while the repulsive interaction destroys neither 
the spin nor charge gaps, the strong attractive interaction
should change the Kondo insulating phase to a new gapless
phase  characterized by
the Luther-Emery model and the $s=1/2$ spin chain model.
In \S 4 the case of the FM Kondo coupling is 
also discussed in detail.
Conclusion is given in \S 5.

\section{Bosonization and scaling equations}

We first introduce the model hamiltonian for the
1D Kondo lattice, and apply
the bosonization method to obtain the scaling equations
which describe the fixed point properties.  
We consider the following hamiltonian,
\begin{equation}
H=-t\sum_{i,\sigma}c^{\dagger}_{\sigma i}c_{\sigma i+1}+h.c.
+U\sum_{i}c^{\dagger}_{\uparrow i}c_{\uparrow i}
c^{\dagger}_{\downarrow i}c_{\downarrow i}
+J_K\sum_{i,\alpha\beta}c^{\dagger}_{\alpha i}\frac{1}{2}
{\mib \sigma}_{\alpha\beta}c_{\beta i}\cdot {\mib S}_{f i},
\label{hamil}
\end{equation}
where $c_{\sigma i}$ ($c^{\dagger}_{\sigma i}$) is an annihilation
(creation) operator for conduction electrons at $i$-th site, 
and ${\mib S}_{f i}$ 
is an operator for localized $ s=1/2$ spins.
The second term is the on-site interaction 
between conduction electrons,
and the last term is the Kondo exchange interaction.
Note that we have  not introduced the RKKY interaction here, 
which was added to the microscopic hamiltonian in our previous paper.
\cite{fuji}
In order to apply renormalization group methods to this model
(\ref{hamil}),
we first bosonize the charge and spin degrees of freedom  of
conduction electrons and take their continuum limit 
with preserving full degrees of 
freedom of localized spins at each lattice site.
Applying abelian\cite{halb} and non-abelian\cite{wit,aff1} 
bosonizations to charge and spin degrees
of freedom of conduction electrons respectively,
we have the effective hamiltonian,
\begin{eqnarray}
H&=&H_c+H_{c-f},  \\
H_c&=& \int {\rm d}x[\frac{v_c}{2K_c}(\partial_x\phi_c(x))^2+
\frac{v_cK_c}{2}\Pi_c^2(x)]
+\frac{U_{um}}{\alpha}\int {\rm d}x
\cos(\sqrt{8\pi}\phi_c(x)+\delta x) \nonumber \\
&+&\int {\rm d}x v_s 
[{\mib J}_{c L}(x)\cdot{\mib J}_{c L}(x)+{\mib J}_{c R}(x)\cdot
{\mib J}_{c R}(x)]-\lambda\int dx {\mib J}_{c L}(x)\cdot{\mib J}_{c R}(x), 
\label{hc}\\
H_{c-f}&=&J_{KA}\int {\rm d}x [{\mib J}_{c L}(x)+{\mib J}_{c R}(x)]
\cdot{\mib S}_f(x) \nonumber \\
&&+J_{KB}\int \frac{{\rm d}x}{\alpha}[
e^{2ik_F x}\frac{1}{2}{\rm tr}(g(x){\mib \sigma})e^{i\sqrt{2\pi}\phi_c(x)}
+h.c.]\cdot {\mib S}_f(x), \label{hcf} 
\end{eqnarray}
where $\phi_c(x)$, $\Pi_c(x)$ are  boson fields for 
the charge degrees of freedom of conduction electrons
and its conjugate momentum, and ${\mib J}_{c L,R}$ is
a spin current operator which satisfies level-1 SU(2) Kac-Moody
algebra, and $g(x)=c^{\dagger}_{\alpha L}c_{\beta R}$ 
($\alpha,\beta=\uparrow,\downarrow$) 
is a fundamental representation of SU(2). 
${\mib S}_{f}(x)=\sum_i f^{\dagger}_{\alpha i}
({\mib \sigma}_{\alpha\beta}/2) f_{\beta i}\delta(x-x_i)$
is the spin density operator for localized spin.
The $U_{um}$- and $\lambda$-terms are, respectively, 
the Umklapp and backward scattering terms of electron-electron 
interaction, and the parameter $\delta$ 
defined by $\delta=4k_{\rm F}-2\pi$ measures the 
deviation from the half-filling case. 
The forward scattering between conduction electrons is incorporated in 
the Tomonaga-Luttinger parameter $K_c$, via 
the initial value $K_c=1/\sqrt{1+2U/\pi v_{\rm F}}$.

In the microscopic model (\ref{hamil}),
the bare parameters should read
 $J_{KA}=J_{KB}=J_K$ and $U_{um}=\lambda=U$, which are
to be modified via the renormalization process.
Note that if we regard the initial 
values of $K_c$, $U_{um}$, and $\lambda$ 
as independent parameters, this effective hamiltonian
describes the low-energy properties of a more general model 
than eq.(\ref{hamil}), which has momentum-dependent 
interactions. We will also extend our arguments 
to such a general model in the subsequent sections.
In what follows we mainly consider the half-filling case $\delta=0$.
Thus we impose the commensurate condition, $2k_{\rm F}a=\pi$
where $a$ is the lattice constant. 

Now we derive the scaling equations for 
various coupling constants of
the effective hamiltonian.
For this purpose we utilize current algebra techniques
for the operators.  The charge field,
$\phi_c$, satisfies the operator product expansion (OPE)
of U(1) Gaussian model,\cite{kada}
\begin{eqnarray}
e^{i\alpha \phi_c(x,\tau)}e^{-i\alpha \phi_c(0,0)}&\sim&
\frac{i\alpha}{\vert z\vert^{\alpha^2K_c/2\pi-2}}
(\frac{\partial_z \phi_c(0,0)}{\bar{z}}
+\frac{\partial_{\bar{z}}\phi_c(0,0)}
{z}) \nonumber \\
+\frac{1}{\vert z\vert^{\alpha^2K_c/2\pi}} 
&-&\frac{\alpha^2}{\vert z\vert^{\alpha^2K_c/2\pi-2}}
\partial_z\phi_c(0,0)
\partial_{\bar{z}}\phi_c(0,0 ) \nonumber \\
-\frac{\alpha^2}{2\vert z\vert^{\alpha^2K_c/2\pi}}
&(&z^2(\partial_z\phi)^2+\bar{z}^2(\partial_{\bar{z}}\phi)^2)
+\cdot\cdot\cdot ,\\
\partial_{z}\phi_c(x,\tau)e^{i\alpha\phi_c(0,0)}&\sim&
\frac{i\alpha K_c}{8\pi z}e^{i\alpha\phi_c(0,0)}+\cdot\cdot\cdot,
\end{eqnarray}
with $z=x+iv_c\tau$, $\bar{z}=x-iv_c\tau$.
${\mib J}_{c L,R}$ and $g(x)$ satisfy the OPE of level-1 SU(2)
Wess-Zumino-Witten model,\cite{kz,bou,pas}
\begin{eqnarray}
J^a_{c L}(x,\tau)J^b_{c L}(0,0)&\sim&\frac{\delta_{ab}}{z^2}
+\frac{\varepsilon^{abc}}
{z}J^c_{c L}(0,0)+\cdot\cdot\cdot, \label{ope1} \\
J^a_{c L}(x,\tau)g(0,0)&\sim&\frac{t^a}{z}g(0,0)+\cdot\cdot\cdot, 
\label{ope2}\\
{\rm tr}[g(x,\tau)\sigma^a]{\rm tr}[g^{\dagger}(0,0)\sigma^b]
&\sim&\frac{\delta_{ab}}{\vert z\vert}+\bigl(\frac{z}{\bar{z}}
\bigr)^{\frac{1}{2}}\varepsilon^{abc}J_{c L}^c(0,0)+\bigl(\frac{\bar{z}}{z}
\bigr)^{\frac{1}{2}}\varepsilon^{abc}J_{c R}^c(0,0) \nonumber \\
&&+\frac{\delta_{ab}}{\vert z \vert}
[z^2J_{c L}^a(0,0)J_{c L}^a(0,0)+\bar{z}^2J_{c R}^a(0,0)J_{c R}^a(0,0)]
+\cdot\cdot\cdot. \label{ope3}
\end{eqnarray}
with $z=x+iv_s\tau$, $\bar{z}=x-iv_s\tau$, and $t^a$, 
the generator of the SU(2) Lie algebra. 
We omit those terms which 
are irrelevant to the following arguments.
Using these formulae, we can derive scaling equations,
\begin{eqnarray}
\frac{d J_{KA}}{d \ln L}&=&\frac{J_{KA}^2}{\pi v_s}
+\frac{J_{KB}^2}{\pi v_s}f_1(\frac{v_c}{v_s}), \label{sca1}\\
\frac{d J_{KB}}{d \ln L}&=&\bigl(\frac{1-K_c}{2}+\frac{2J_{KA}}{\pi v_s}
+\frac{2\lambda}{\pi v_s}+\frac{2U_{um}}{\pi v_c}\bigr)J_{KB}, \label{sca2}\\
\frac{d \lambda}{d \ln L}&=&-\frac{\lambda^2}{\pi v_s}, \label{sca3}\\
\frac{d U_{um}}{d \ln L}&=&(2-2K_c)U_{um}, \label{sca4}\\
\frac{d K_c}{d \ln L}&=&-K_c^2\frac{U^2_{um}}{v_c^2}
-K_c^2\frac{J_{KB}^2}{v_s^2}
f_2(\frac{v_c}{v_s}), \label{sca5} 
\end{eqnarray}
with
\begin{eqnarray}
f_1(u)&=&\int^{\pi}_{0}d \theta\frac{\sin\theta}{(\cos^2\theta+u^2\sin^2\theta
)^{\frac{K_c}{2}}}, \\
f_2(u)&=&\int^{\pi}_0 d\theta\frac{1}{(\cos^2\theta+u^2\sin^2\theta
)^{\frac{K_c}{2}-1}}.
\end{eqnarray}
In the dilute limit of localized spins,  
the velocity of conduction electrons is not renormalized.
In this case, if one neglects backward and Umklapp processes 
of electron-electron interaction, and expand the Luttinger 
liquid parameter up to the lowest order in the forward scattering, 
$K_c\sim 1-U/\pi$,
one can reproduce the scaling equations for the single impurity Kondo
problem in Luttinger liquids, which were obtained by 
g-ology methods.\cite{furu}
In the lattice system, the RKKY interaction between localized spins
is generated by higher order processes of 
the Kondo exchange interaction. 
With the use of eq.(\ref{ope3}), we obtain the RKKY interaction
from the second order of $J_{KB}$,
\begin{equation}
H_{RKKY}=-J_{KB}^2\int dx\int dx'f_3(\frac{v_c}{v_s})
\frac{\cos2k_{\rm F}(x-x')}{\vert x-x'\vert}
{\mib S}_f(x)\cdot{\mib S}_f(x'),
\label{rkky}
\end{equation}
with
\begin{equation}
f_3(u)=\int^{\pi}_0 {\rm d}\theta 
\frac{1}{(\cos^2\theta+u^2\sin^2\theta)^{\frac{K_c}{2}}}.
\end{equation} 
The predominant contribution comes from the nearest neighbor interaction.
Since $2k_{\rm F}a=\pi$, the RKKY interaction is antiferromagnetic 
for nearest neighbor sites.
Thus we replace the RKKY interaction term approximately by
\begin{equation}
H_{RKKY}=J_{KB}^2 \ln L f_3(\frac{v_c}{v_s})\sum_i {\mib S}_{f i}\cdot
{\mib S}_{f i+1}.
\end{equation}
Then the scaling equation for the RKKY interaction is given by
\begin{equation} 
\frac{d J_{RKKY}}{d \ln L}=\frac{J_{KB}^2}{\pi v_s}f_3(\frac{v_c}{v_s}).
\label{sca6}
\end{equation}
Eqs.(\ref{sca1})$\sim$(\ref{sca5}) and (\ref{sca6}) determine 
the low-energy properties of the model.

We note here that in our previous paper the RKKY
interaction was naively introduced in the 
microscopic hamiltonian.
\cite{fuji} In the following we will take into account 
that the RKKY term is a higher-order term generated
by the Kondo interaction, and see what follows from
this  fact.

\section{Antiferromagnetic Kondo coupling case}

We start our discussions with the case of AFM Kondo coupling
($J_K>0$).
In this case, from eqs.({\ref{sca1}}) and (\ref{sca6}),
we see that $J_{KA}$ and $J_{RKKY}$ flow to the strong-coupling 
regime. Since initially $J_K>J_{RKKY}=0$, the Kondo interaction $J_K$
is predominant over the RKKY interaction $J_{RKKY}$.
Thus low-energy properties at the fixed point 
are determined essentially by  $J_K$.
We have three different situations according to the interaction 
strength for conduction electrons: {\it case A}: $U=0$ and $K_c(0)=1$,
{\it case B}: 
$U>0$ and $K_c(0)<1$, and {\it case C}: $U<0$ and $K_c(0)>1$. 

\subsection{case A: $U=0$ and $K_c(0)=1$}

In the absence of interaction between conduction electrons, $U=0$,
the scaling equations for the Kondo couplings are simplified as,
\begin{eqnarray}
\frac{d J_{KA}}{d \ln L}&=&\frac{J_{KA}^2}{\pi v_s}+\frac{J_{KB}^2}{\pi v_s}
f_1(\frac{v_c}{v_s}), \\
\frac{d J_{KB}}{d \ln L}&=&\frac{2J_{KA}J_{KB}}{\pi v_s}.
\end{eqnarray}
Thus $J_{KA}$ and $J_{KB}$ are marginally relevant, and produce
spin and charge gaps.
The fixed point is the Kondo insulator 
as pointed out before.\cite{tsu,fuji} 
If one neglects the renormalization of the velocity, the 
spin gap $\Delta_s$ and the charge gap $\Delta_c$
generated by $J_{KA}$ and $J_{KB}$ are given by
\begin{equation}
\Delta_{s, c} \sim e^{-\frac{\pi v_s}{J_K}},
\end{equation}
for $J_K\ll v_s$. The $J_K$-dependence of the spin 
gap coincides with that obtained by Tsunetsugu et al.\cite{tsu}.

\subsection{case B: $U>0$ and $K_c(0)<1$}

When  the on-site interaction for conduction electrons
is repulsive, the resulting 
$\lambda$-term (backward scattering)
is marginally irrelevant, and negligible.
$J_{KB}$ is a relevant interaction with scaling dimension
$3/2+K_c/2$ as seen from eq.(\ref{sca2}).
Since this dimension is larger than that of the Umklapp interaction,
the Umklapp interaction $U_{um}$ 
grows faster than $J_{KB}$ at half-filling, which may 
give rise to the Mott-Hubbard gap 
in the charge sector of conduction electrons.
 However if the initial value of $U_{um}$ is 
much smaller than that of $J_{KB}$,
$J_{KB}$-term may generate the charge gap before $U_{um}$-term 
flows into the strong-coupling regime.
Thus several different cross-over behaviors are expected to
appear according to the initial values of
parameters. In the following we show the 
expressions for spin and charge gaps in three limiting cases.

\vskip3mm
\noindent
(i) {\it $\frac{1-K_c(0)}{2}\gg \frac{2J_{KA}(0)}{\pi v_s(0)}$ 
and $U_{um}(0)/J_K(0)\gtrsim 1$.}
\vskip2mm

The Mott-Hubbard gap due to $U_{um}$-term opens prior to 
the spin gap formation due to $J_K$ in this case. 
After the charge gap formation, $K_c=0$, and eq.(\ref{sca2}) becomes,
\begin{equation}
\frac{d J_{KB}}{d \ln L}=\frac{J_{KB}}{2}
\end{equation}
Thus the spin gap for small $J_K$ is given by
\begin{equation}
\Delta_s\sim J_K^2. \label{sgp1}
\end{equation}
The charge gap is given by the Mott-Hubbard gap of the 1D Hubbard model, 
if $J_K$ is sufficiently small compared to $U$.

\vskip3mm
\noindent
(ii) {\it $\frac{1-K_c(0)}{2}\gg \frac{2J_{KA}(0)}{\pi v_s(0)}$ and 
 $U_{um}(0)/J_K(0)\ll 1$.}
\vskip2mm

In this case, the charge gap generation due to $J_{KB}$-term 
is predominant over the Mott transition due to 
the Umklapp interaction between conduction electrons.
Thus the spin and charge gap is given by
\begin{equation}
\Delta_{s,c}\sim J_K^{\frac{2}{1-K_c}}, \label{sgp2}
\end{equation}
for small $J_K$.
The realization of this case requires a
large forward scattering of electron-electron interaction
and a small Umklapp interaction at the initial stage 
of the renormalization process.
However, in the microscopic model given by eq.(\ref{hamil}),
these parameters are initially equal to each other.
Thus in order to actually observe this behavior, we should consider 
a more general model such as the g-ology model which has 
different coupling constants of 
electron-electron interaction 
depending on transfered momentum.\cite{sol} 

\vskip3mm
\noindent
(iii) {\it $\frac{1-K_c(0)}{2}\ll \frac{2J_{KA}(0)}{\pi v_s(0)}$.}
\vskip2mm

In this case, the first term of the right hand side of eq.(\ref{sca2})
is negligible. Thus the spin gap has the same $J_K$-dependence 
as the case of $K_c=1$ ($U=0$ case).
\begin{equation}
\Delta_s\sim e^{-\frac{\pi v_s}{J_K}}.
\end{equation}
The charge gap in this case is generated 
by $U_{um}$-term or $J_{KB}$-term
depending on the initial value of these couplings. 

It is to be noted here  that the above different behaviors 
observed in the gap formation do not imply 
that there appear new different universality classes: 
the system still belongs to the same universality  of
the Kondo insulating phase as far as $U \ge 0$, 
and  cross-over behaviors rather than 
phase boundaries are seen when  
the initial values of bare parameters are changed.

We would like to compare the  above results with those obtained by 
Shibata et al.\cite{shi}.
They calculated numerically the spin gap of the model 
given by eq.(\ref{hamil}).
They concluded that the spin gap is given by
$\Delta_s\sim \exp(-\pi v_s/\alpha J_K)$ for small $J_K$ and any $U$, 
where a parameter $\alpha$ depends on $U$.
Their conclusion does not seem to coincide with our results 
in the case (i) shown above.
Actually, the parameter region where their calculation was performed
corresponds to the case (iii) or the case where $(1-K_c(0))/2$ is 
of the same order as $J_{KA}(0)/\pi v_s(0)$.
We have solved numerically the scaling equations 
when $(1-K_c(0))/2$ is 
of the same order as $J_{KA}(0)/\pi v_s(0)$, and 
deduced the functional dependence of the spin gap on $J_K$. 
We found that the results  fit in with the
exponential function fairly well.
Thus our results are consistent with those of Shibata et al.
We think that the value of $J_K$ they used may not be small enough 
to observe the power-law dependence of eq.(\ref{sgp1}).

\subsection{case C: $U<0$ and $K_c(0)>1$}

In the attractive case of $U$, the Umklapp term 
is irrelevant, and can be discarded.
However since $\lambda<0$, the $\lambda$-term is marginally relevant.
The $J_{KB}$-term may be relevant or irrelevant 
depending on $K_c(0)$ and $J_K$. Therefore 
we have three different behaviors as shown below.

\vskip3mm
\noindent
(i) {\it $J_{KB}$ predominates over $J_{KA}$ and $\lambda$.}
\vskip2mm

If $\vert U\vert$ is sufficiently small compared to $J_K$, 
and $K_c(0)$ is not so large,
the $J_{KB}$-term is still relevant.
The spin and charge gaps are generated by the $J_{KA}$- and $J_{KB}$-terms. 
For small $J_K$, the gaps are given by
\begin{equation}
\Delta_{s, c}\sim e^{-\frac{\pi v_{s}}{J_K}}.
\end{equation}
Note that the universality class still belongs to that
of the Kondo insulator.

\vskip3mm
\noindent
(ii) {\it $J_{KA}$ predominates over $J_{KB}$ and $\lambda$.}
\vskip2mm

If $K_c(0)$ is sufficiently large, the $J_{KB}$-term becomes irrelevant.
Moreover if $\vert\lambda(0)\vert$ is smaller than $J_K$,
$J_{KA}$ grows to a strong-coupling
value before $\lambda$ flows into the strong-coupling regime. 
In this case, the $J_{KA}$-term generates the spin gap given by
\begin{equation}
\Delta_s\sim e^{-\frac{\pi v_s}{J_K}}.
\end{equation}
The charge sector is massless.
Thus the fixed point of the charge degrees of freedom 
is the Tomonaga-Luttinger 
liquids described by $c=1$ U(1) Gaussian model.
This would bring about a new phase driven by the interaction.
However, we recall that for the microscopic 
model given by eq.(\ref{hamil}), 
we can not control the parameters 
$K_c(0)$ and $\lambda(0)$ independently.
Thus it is not obvious whether this phase may
realize for the model (\ref{hamil}). If one considers 
the model in which the strength of the forward scattering of 
electron-electron interaction which controls $K_c(0)$ 
and the backward scattering $\lambda(0)$ 
are independent parameters, one can surely find this phase.

\vskip3mm
\noindent
(iii) {\it $\lambda$ predominates over $J_{KA}$ and $J_{KB}$.}
\vskip2mm

We now discuss the case where $\vert U\vert$ is sufficiently large 
and $\vert U\vert \gg J_K$.
Then the $J_{KB}$-term is irrelevant,
and thus the charge sector becomes  massless,
forming the Tomonaga-Luttinger liquid described by
$c=1$ U(1) Gaussian model. 
On the other hand, the spin gap generation occurs 
due to the $\lambda$-term (backward scattering term) 
prior to that due to the $J_{KA}$-term. 
This state obtained for conduction electrons
is nothing but a metallic state with the spin gap
which is often referred to as the Luther-Emery class.
Localized spins are now completely decoupled from 
conduction electrons, making a massless mode classified  
as SU(2) Tomonaga-Luttinger liquid described by 
$c=1$ level-1 SU(2) Wess-Zumino-Witten model.
Therefore it turns out that a sufficiently large 
attractive interaction should drive the Kondo insulator to
a completely different phase, in which both of the
spin and charge sectors are massless: 
conduction electrons are in the Luther-Emery phase and  
localized spins are in  SU(2) Tomonaga-Luttinger liquid.
Recently Shibata et al. have also found this massless phase 
for large $\vert U\vert$
by using the density matrix renormalizatoin group method.\cite{shi2}

\section{Ferromagnetic Kondo coupling case}

We now move to the discussions for the case of FM Kondo
couplings. It is known that the spin gap a la Haldane
opens at half filling in this case.\cite{tsu,hal}
 We now explore how the formation 
of the spin and charge  gaps is affected by the electron-electron
interaction for conduction electrons.

\subsection{$U=0$ and $K_c(0)=1$}

In this case, as is seen from eqs.(\ref{sca1}) and (\ref{sca2}),
the $J_{KA}$- and $J_{KB}$-terms are marginally irrelevant, and 
these couplings scale to zero.
In the process of the renormalization, $J_{RKKY}$ becomes larger
than $\vert J_{KA}\vert$ and $\vert J_{KB}\vert$.
Thus we can not discard the RKKY interaction 
compared to the Kondo coupling.  
The induced RKKY interaction gives rise 
to a dispersion of localized spins, 
and thus changes the scaling equations obtained in \S 2.
This point is quite contrasted to
the antiferromagnetic case discussed above, where 
the $J_{KA}$- and $J_{KB}$-terms are marginally relevant.
For $J_{RKKY}\gg \vert J_{KA}\vert, \vert J_{KB}\vert$,
we can bosonize localized spins in the continuum limit, and treat 
the Kondo coupling perturbatively.
Then the field theoretical limit of the RKKY interaction is
\begin{equation}
H_{RKKY}=\int {\rm d}x v_{f}[{\mib J}_{f L}(x)\cdot{\mib J}_{f L}(x)+
{\mib J}_{f R}(x)\cdot{\mib J}_{f R}(x)]
-\lambda_f\int {\rm d}x{\mib J}_{f L}(x)\cdot
{\mib J}_{f R}(x),
\end{equation}
where ${\mib J}_{f L, R}$ satisfy level-1 SU(2) Kac-Moody algebra.
$\lambda_f>0$ is a coupling constant for the marginally 
irrelevant operator
which exists because of chiral SU(2)$\times$SU(2) symmetry. 
This is level-1 SU(2) Wess-Zumino-Witten model with marginally 
irrelevant interaction.
We now incorporate this RKKY term into 
 a non-perturbed part of the hamiltonian.
Then the bosonized expression of the Kondo exchange interaction 
is altered to
\begin{eqnarray}
H_{c-f}&=&J_{KA}\int {\rm d}x ({\mib J}_{c L}(x)+{\mib J}_{c R}(x))\cdot
({\mib J}_{f L}(x)+{\mib J}_{f R}(x)) \nonumber \\
&& +J_{KB} \int {\rm d}x [{\rm tr}(g(x)\frac{\mib \sigma}{2})\cdot{\rm tr}
(g_f(x)\frac{\mib \sigma}{2})e^{i\sqrt{2\pi}\phi_c(x)}+h.c.],
\label{kf}
\end{eqnarray}
where $g_f(x)=f^{\dagger}_{\alpha L}(x)f_{\beta R}(x)$ 
($\alpha,\beta=\uparrow,\downarrow$).
The scaling dimension of the second term of eq.(\ref{kf})  
is now $1+K_c/2$.
Note that $K_c$ is renormalized by the $J_{KB}$-term and 
should be smaller than 1.
This term is relevant and generates the spin and charge gaps.
The low-energy fixed point is nothing but 
the Haldane gap phase.\cite{tsu} 
Although the spin and charge gaps are given by 
the power-law function of $J_{KB}$, 
we cannot write down their expressions explicitly in terms of
bare coupling $J_K$, because $J_{KB}$ in eq.(\ref{kf}) 
is renormalized, being different from bare $J_K$. 

\subsection{$U>0$ and $K_c(0)<1$}

If we introduce the repulsive 
interaction for conduction electrons,
 $J_{KB}$ becomes relevant, as is seen from eq.(\ref{sca2}).
Thus the spin and charge gaps open  
as discussed in \S {\it 3.2}.
The spin gap is given by eq.(\ref{sgp1}) or (\ref{sgp2}).
The universality class is still characterized by
the insulating phase with the Haldane-gap.

\subsection{$U<0$ and $K_c(0)>1$}  

In the case of attractive interaction, 
the $J_{KA}$- and $J_{KB}$-terms of eq.(\ref{hcf}) 
and the Umklapp interaction $U_{um}$ are irrelevant.
Thus we should take into account the RKKY interaction 
again as in the case of $U=0$.
We bosonize localized spin and use eq.(\ref{kf}) as the effective Kondo 
exchange interaction.
Then the scaling equation for $J_{KB}$ is changed to
\begin{equation}
\frac{d J_{KB}}{d \ln L}=\bigl(1-\frac{K_c}{2}+\frac{2J_{KA}}{\pi v_s}
+\frac{2\lambda}{\pi v_s}
+\frac{2\lambda_f}{\pi v_f}+\frac{2U_{um}}{\pi v_c}\bigr)J_{KB}. \label{scafn}
\end{equation}
The $J_{KB}$-term may be relevant or irrelevant depending on the initial 
values of $\lambda=\lambda_0$ and $K_c=K_{c 0}$,
which are renormalized to the values different from $U$ and $K_c(0)$. 
Moreover, as is discussed in \S {\it 3.3}, 
the $\lambda$-term flows to a strong-coupling regime.
Thus we have two possibilities: 
(i) $J_{KB}$ predominates over $\lambda$, 
(ii) $\lambda$ predominates over $J_{KB}$.

\vskip3mm
\noindent
(i) {\it $J_{KB}$ predominates over $\lambda$.}
\vskip2mm

If $\vert\lambda_0\vert$ is sufficiently small, and $K_{c 0}$ 
is not so large, $J_{KB}$ may be relevant.
Since $J_{KB}$ grows faster than $\lambda$, the $J_{KB}$-term generates
spin and charge gaps prior to the spin gap formation due to 
the $\lambda$-term provided 
that the initial value of $J_{KB}$ is not much smaller than
$\lambda_0$. The fixed point is the same class as the Haldane gap phase.

\vskip3mm
\noindent
(ii) {\it $\lambda$ predominates over $J_{KB}$.}
\vskip2mm

For sufficiently large negative $U$,
$\lambda$ flows to a strong-coupling regime, $\lambda\rightarrow -\infty$, 
faster than $J_{KB}$. Then, the right hand side of eq.(\ref{scafn})
becomes positive. The $J_{KB}$-term is irrelevant in this case. 
Thus the charge sector of conduction electrons
is massless, and described by $c=1$  U(1) Gaussian model, i.e.
the Tomonaga-Luttinger liquids, and 
the spin gap of conduction electrons opens
due to the $\lambda$-term, as has been discussed for the
AFM case.  Therefore, even for the FM case,
a sufficiently strong attractive interaction 
brings the system to a massless phase with
conduction electrons in the Luther-Emery phase
and localized spins in the SU(2) Tomonaga-Luttinger phase.


\section{Conclusion}

We have studied the 1D Kondo lattice model 
by means of bosonization techniques, and discussed 
how the mass gap is generated at half filling.  
It has been confirmed that
the spin and charge gaps are indeed generated 
both for the AFM and FM Kondo couplings,
realizing the Kondo insulating phase 
and the Haldane-gap insulating phase.
By taking into account that the RKKY 
interaction is a higher order term induced 
by the Kondo coupling, 
we have shown that the generation of the spin gap occurs
in different manners for the AFM and FM cases. 
This is one of the central points 
to be claimed in  the 
present paper, which was not considered in our  previous paper.\cite{fuji}
The results thus obtained for the spin gap are consistent 
with the previous numerical studies, and may explain
why the functional dependence of the spin gap
on $J_K$ can differ between the AFM and FM couplings.\cite{tsu} 

We have also investigated how the  
short-range interaction between conduction 
electrons affects on the formation of these excitation gaps.
It has been shown that although the repulsive interaction 
changes the functional dependence 
of spin and charge gaps on the Kondo coupling, 
the universality class is still characterized by that for $U=0$. 
On the other hand, a sufficiently strong attractive interaction can  
destroy the spin and charge gaps,
resulting in a completely different class of  
massless liquids consisting of 
the  Luther-Emery state and  
the SU(2) Tomonaga-Luttinger liquid state.

We have seen that the spin gap formation 
can be described rather well in the present 
approach. As for the charge gap, however, we have not been 
able to completely resolve
the discrepancy between the present results and the 
numerical results, although the qualitative agreement was 
already established concerning whether 
the mass is generated or not. 
For example, the exponential dependence of the charge gap 
on $J_K$ obtained for the AFM case with $U=0$ does not 
seem to fit in with the $J_K$-linear dependence deduced 
numerically.\cite{ueda} To resolve this problem 
is one of the main issues to be addressed in the future work.

Finally we wish to note that
the  present results are applicable 
not only to the model with the Hubbard-like
on-site interaction but also to more general models with
momentum-dependent electron-electron interaction. 
For such general models, one can find 
the metallic phase with spin gap predicted in \S 3
for the AFM Kondo model with attractive 
electron-electron interaction.

\acknowledgement
One of the authors (S.F.) thanks N. Shibata for 
valuable discussions.
This work was partly supported by a Grant-in-Aid from the Ministry
of Education, Science and Culture.
  


\end{document}